\documentclass[%
 reprint,
superscriptaddress,
nofootinbib,
 amsmath,amssymb,
 aps,
 prl,
]{revtex4-1}
\usepackage{multirow}
\usepackage{graphicx}
\usepackage{siunitx}
\usepackage{xcolor}
\usepackage{subcaption}
\usepackage[utf8x]{inputenc}
\usepackage{ulem}
\usepackage{supertabular}
\usepackage{hyperref}

\definecolor{gesfblack}{rgb}{0,0,0}

\definecolor{gesfblue}{rgb}{0.08,0.42,0.76}

\definecolor{gesfgreen}{rgb}{0,1,0}

\definecolor{gesfgrey}{rgb}{0.5,0.5,0.5}

\definecolor{gesflanse}{rgb}{0.00,0.50,0.50}

\definecolor{gesfpurple}{rgb}{0.47,0.19,0.42}

\definecolor{gesfred}{rgb}{1,0,0}

\definecolor{gesfwhite}{rgb}{1,1,1}

\definecolor{gesfyellow}{rgb}{0.7,0.4,0.3}

\begin{document}

\title{A Search for Light Fermionic Dark Matter Absorption on Electrons in PandaX-4T}

\begin{abstract}
We report a search on a sub-MeV fermionic dark matter 
absorbed by electrons with an outgoing active neutrino using the 0.63~tonne$\cdot$year exposure collected by PandaX-4T liquid xenon experiment. 
No significant signals are observed over the expected background.
The data are interpreted into limits to the effective couplings between such dark matter and the electron. For axial-vector or vector interactions, our sensitivity is competitive in comparison to existing astrophysical bounds on the decay of such a dark matter candidate into photon final states. In particular, we present the first direct detection limits for a vector~(axial-vector) interaction which are the strongest in the mass range from 35 to 55~(25 to 45)~keV/c$^2$. 

\end{abstract}

\def\shKeyLab{School of Physics and Astronomy, Shanghai Jiao Tong University, Key Laboratory for Particle Astrophysics and Cosmology (MoE), Shanghai Key Laboratory for Particle Physics and Cosmology, Shanghai 200240, China}
\def\BUAA{School of Physics, Beihang University, Beijing 102206, China}
\def\BUAALab{Beijing Key Laboratory of Advanced Nuclear Materials and Physics, Beihang University, Beijing, 102206, China}
\def\zzu{School of Physics and Microelectronics, Zhengzhou University, Zhengzhou, Henan 450001, China}
\def\USTClab{State Key Laboratory of Particle Detection and Electronics, University of Science and Technology of China, Hefei 230026, China}
\def\USTCdep{Department of Modern Physics, University of Science and Technology of China, Hefei 230026, China}
\def\BUAALab{International Research Center for Nuclei and Particles in the Cosmos \& Beijing Key Laboratory of Advanced Nuclear Materials and Physics, Beihang University, Beijing 100191, China}
\def\pku{School of Physics, Peking University, Beijing 100871, China}
\def\YaLongSD{Yalong River Hydropower Development Company, Ltd., 288 Shuanglin Road, Chengdu 610051, China}
\def\IAP{Shanghai Institute of Applied Physics, Chinese Academy of Sciences, 201800 Shanghai, China}
\def\CHEPpku{Center for High Energy Physics, Peking University, Beijing 100871, China}
\def\SDUdep{Research Center for Particle Science and Technology, Institute of Frontier and Interdisciplinary Scienc, Shandong University, Qingdao 266237, Shandong, China}
\def\SDUlab{Key Laboratory of Particle Physics and Particle Irradiation of Ministry of Education, Shandong University, Qingdao 266237, Shandong, China}
\def\UMD{Department of Physics, University of Maryland, College Park, Maryland 20742, USA}
\def\TDLee{Tsung-Dao Lee Institute, Shanghai Jiao Tong University, Shanghai, 200240, China}
\def\MESJTU{School of Mechanical Engineering, Shanghai Jiao Tong University, Shanghai 200240, China}
\def\SYU{School of Physics, Sun Yat-Sen University, Guangzhou 510275, China}
\def\SYUSFI{Sino-French Institute of Nuclear Engineering and Technology, Sun Yat-Sen University, Zhuhai, 519082, China}
\def\NKU{School of Physics, Nankai University, Tianjin 300071, China}
\def\FDU{Key Laboratory of Nuclear Physics and Ion-beam Application (MOE), Institute of Modern Physics, Fudan University, Shanghai 200433, China}
\def\USST{School of Medical Instrument and Food Engineering, University of Shanghai for Science and Technology, Shanghai 200093, China}
\def\SJTUSC{Shanghai Jiao Tong University Sichuan Research Institute, Chengdu 610213, China}
\def\Princeton{Physics Department, Princeton University, Princeton, NJ 08544, USA}
\def\MIT{Department of Physics, Massachusetts Institute of Technology, Cambridge, MA 02139, USA}
\def\SARI{Shanghai Advanced Research Institute, Chinese Academy of Sciences, Shanghai 201210, China}
\def\SPEIT{SJTU Paris Elite Institute of Technology, Shanghai Jiao Tong University, Shanghai, 200240, China}
\def\taiWan{Department of Physics, National Taiwan University, Taipei 10617}

\affiliation{\shKeyLab}
\author{Dan Zhang}\email[Corresponding author: ]{dzhang16@umd.edu}\affiliation{\UMD}
\author{Abdusalam Abdukerim}\affiliation{\shKeyLab}
\author{Zihao Bo}\affiliation{\shKeyLab}
\author{Wei Chen}\affiliation{\shKeyLab}
\author{Xun Chen}\affiliation{\shKeyLab}\affiliation{\SJTUSC}
\author{Yunhua Chen}\affiliation{\YaLongSD}
\author{Chen Cheng}\affiliation{\SYU}
\author{Zhaokan Cheng}\affiliation{\SYUSFI}
\author{Xiangyi Cui}\affiliation{\TDLee}
\author{Yingjie Fan}\affiliation{\NKU}
\author{Deqing Fang}\affiliation{\FDU}
\author{Changbo Fu}\affiliation{\FDU}
\author{Mengting Fu}\affiliation{\pku}
\author{Lisheng Geng}\affiliation{\BUAA}\affiliation{\BUAALab}\affiliation{\zzu}
\author{Karl Giboni}\affiliation{\shKeyLab}
\author{Linhui Gu}\affiliation{\shKeyLab}
\author{Xuyuan Guo}\affiliation{\YaLongSD}
\author{Ke Han}\affiliation{\shKeyLab}
\author{Changda He}\affiliation{\shKeyLab}
\author{Jinrong He}\affiliation{\YaLongSD}
\author{Di Huang}\affiliation{\shKeyLab}
\author{Yanlin Huang}\affiliation{\USST}
\author{Zhou Huang}\affiliation{\shKeyLab}
\author{Ruquan Hou}\affiliation{\SJTUSC}
\author{Xiangdong Ji}\affiliation{\UMD}
\author{Yonglin Ju}\affiliation{\MESJTU}
\author{Chenxiang Li}\affiliation{\shKeyLab}
\author{Jiafu Li}\affiliation{\SYU}
\author{Mingchuan Li}\affiliation{\YaLongSD}
\author{Shu Li}\affiliation{\MESJTU}
\author{Shuaijie Li}\affiliation{\TDLee}
\author{Qing Lin}\affiliation{\USTClab}\affiliation{\USTCdep}
\author{Jianglai Liu}\email[Spokesperson: ]{jianglai.liu@sjtu.edu.cn}\affiliation{\shKeyLab}\affiliation{\TDLee}\affiliation{\SJTUSC}
\author{Xiaoying Lu}\affiliation{\SDUdep}\affiliation{\SDUlab}
\author{Lingyin Luo}\affiliation{\pku}
\author{Yunyang Luo}\affiliation{\USTCdep}
\author{Wenbo Ma}\affiliation{\shKeyLab}
\author{Yugang Ma}\affiliation{\FDU}
\author{Yajun Mao}\affiliation{\pku}
\author{Nasir Shaheed}\affiliation{\SDUdep}\affiliation{\SDUlab}
\author{Yue Meng}\affiliation{\shKeyLab}\affiliation{\SJTUSC}
\author{Xuyang Ning}\affiliation{\shKeyLab}
\author{Ningchun Qi}\affiliation{\YaLongSD}
\author{Zhicheng Qian}\affiliation{\shKeyLab}
\author{Xiangxiang Ren}\affiliation{\SDUdep}\affiliation{\SDUlab}
\author{Changsong Shang}\affiliation{\YaLongSD}
\author{Xiaofeng Shang}\affiliation{\shKeyLab}
\author{Guofang Shen}\affiliation{\BUAA}
\author{Lin Si}\affiliation{\shKeyLab}
\author{Wenliang Sun}\affiliation{\YaLongSD}
\author{Andi Tan}\affiliation{\UMD}
\author{Yi Tao}\affiliation{\shKeyLab}\affiliation{\SJTUSC}
\author{Anqing Wang}\affiliation{\SDUdep}\affiliation{\SDUlab}
\author{Meng Wang}\affiliation{\SDUdep}\affiliation{\SDUlab}
\author{Qiuhong Wang}\affiliation{\FDU}
\author{Shaobo Wang}\affiliation{\shKeyLab}\affiliation{\SPEIT}
\author{Siguang Wang}\affiliation{\pku}
\author{Wei Wang}\affiliation{\SYUSFI}\affiliation{\SYU}
\author{Xiuli Wang}\affiliation{\MESJTU}
\author{Zhou Wang}\affiliation{\shKeyLab}\affiliation{\SJTUSC}\affiliation{\TDLee}
\author{Yuehuan Wei}\affiliation{\SYUSFI}
\author{Mengmeng Wu}\affiliation{\SYU}
\author{Weihao Wu}\affiliation{\shKeyLab}
\author{Jingkai Xia}\affiliation{\shKeyLab}
\author{Mengjiao Xiao}\affiliation{\UMD}
\author{Xiang Xiao}\affiliation{\SYU}
\author{Pengwei Xie}\affiliation{\TDLee}
\author{Binbin Yan}\affiliation{\shKeyLab}
\author{Xiyu Yan}\affiliation{\USST}
\author{Jijun Yang}\affiliation{\shKeyLab}
\author{Yong Yang}\affiliation{\shKeyLab}
\author{Chunxu Yu}\affiliation{\NKU}
\author{Jumin Yuan}\affiliation{\SDUdep}\affiliation{\SDUlab}
\author{Ying Yuan}\affiliation{\shKeyLab}
\author{Xinning Zeng}\affiliation{\shKeyLab}
\author{Minzhen Zhang}\affiliation{\shKeyLab}
\author{Peng Zhang}\affiliation{\YaLongSD}
\author{Shibo Zhang}\affiliation{\shKeyLab}
\author{Shu Zhang}\affiliation{\SYU}
\author{Tao Zhang}\affiliation{\shKeyLab}
\author{Yuanyuan Zhang}\affiliation{\TDLee}
\author{Li Zhao}\affiliation{\shKeyLab}
\author{Qibin Zheng}\affiliation{\USST}
\author{Jifang Zhou}\affiliation{\YaLongSD}
\author{Ning Zhou}\affiliation{\shKeyLab}
\author{Xiaopeng Zhou}\affiliation{\BUAA}
\author{Yong Zhou}\affiliation{\YaLongSD}
\author{Yubo Zhou}\affiliation{\shKeyLab}

\collaboration{PandaX Collaboration}

\author{Shao-Feng Ge}\email[Corresponding author: ]{gesf@sjtu.edu.cn}\affiliation{\TDLee}\affiliation{\shKeyLab}
\noaffiliation
\author{Xiao-Gang He}\affiliation{\TDLee}\affiliation{\taiWan}
\author{Xiao-Dong Ma}\affiliation{\TDLee}\affiliation{\shKeyLab}
\author{Jie Sheng}\affiliation{\TDLee}\affiliation{\shKeyLab}
\maketitle

\section{Introduction}
The nature of dark matter~(DM), which makes up around 85\% of the total mass in the Universe~\cite{planck}, is a top scientific mystery. 
The cold dark matter featured in the standard cosmological model ($\Lambda$CDM) is consistent with large scale cosmological observations at different eras of the cosmic evolution.
Leading cold dark matter candidates include the Weakly Interacting Massive Particles (WIMPs) and axions, which have been searched with tailored experiments for decades~\cite{p4t,lux,xenon_nt,lzprojected,xenon_1t,darkside50,darkside20k,pico,picasso,admx,haystac,capp}.
Warm dark matter, on the other hand, is also a viable dark matter solution, which mitigates the so-called small scale problems of $\Lambda$CDM~\cite{coreCusp,coreCusp2,missingSatellite,missingSatellite2,tooBigToFail} with a larger free streaming distance~\cite{wdm,wdm1,dw}, while maintaining excellent agreement with observations at the large scale.
A representative warm dark matter candidate is the keV-scale sterile neutrino~\cite{keVnuWP}, which received particular attention recently due to the 3.5~keV unidentified excess in the x-ray spectrum from satellites~\cite{3.55line,3.55line2,3.55line3}.
However, comprehensive analyses later have challenged the sterile neutrino interpretations of the excess~\cite{nuIncon2,nuIncon}.

Nevertheless, a more general light neutral fermionic particle can still be a warm or cold DM candidate, depending on the initial conditions in the thermal history. 
If such fermionic particles (noted as $\chi$ hereafter) couple to neutrinos or charged leptons via an effective interaction, they can be probed experimentally~\cite{absorptionMain,ytt,gsf}.
In a DM direct detection experiment, the tree-level process of $\chi$ absorbed on an electron with an outgoing active neutrino ($\chi e\to e\nu$) can be sensitively searched, as the electron obtains a kinetic recoil through the mass of $\chi$~\cite{ytt}. Such a search becomes particularly attractive for axial-vector~(A) and vector~(V) interactions, where the satellite x-ray limits are relatively weaker since $\chi\to\nu\gamma$ can only be produced by two-loop diagrams due to the gauge symmetry and quantum electrodynamics charge conjugation symmetry~\cite{gsf}. 
In fact, the recent electron recoil excess observed in XENON1T was interpreted as a fermionic DM with a mass of 60~keV/c$^2$ absorbed on electrons via a vector interaction~\cite{ytt,gsf,liTong}. However, such interpretation  {has some tension with cosmological observations}  as the invisible decay of $\chi\to3\nu$
would have injected more radiation into the early universe, leaving traces in the Hubble constant and matter power spectrum~\cite{gsf}. For these DM candidates, constraints 
also arise from the relic density to avoid overproduction through the ``freeze-in" mechanism~\cite{cosLimit}.
In this letter, we present the first dedicated search on fermionic DM-electron absorption using recent data from PandaX-4T experiment {to further elucidate the situation.}


For the effective interaction, the V and A operators can be written as~\cite{gsf}
\begin{equation}\label{eq:operator}
\begin{split}
     &\mathcal{O}^{V}_{e\nu\chi}=\frac{1}{\Lambda^2}(\bar{e}\gamma_{\mu}e)(\bar{\nu}_{L}\gamma^{\mu}\chi_{L})\\
    &\mathcal{O}^{A}_{e\nu\chi}=\frac{1}{\Lambda^2}(\bar{e}\gamma_{\mu}\gamma_{5}e)(\bar{\nu}_{L}\gamma^{\mu}\chi_{L})
\end{split},
\end{equation}
where the standard model~(SM) left-handed neutrino is taken, 
and $1/\Lambda^2$ is the Wilson coefficient with a dimension of inverse mass square. Note that our experimental constraint stays the same if a light sterile neutrino ($m_{\nu} \ll m_{\chi}$) replaces the active neutrino in Eq.~\ref{eq:operator}, and the changes in other astrophysical and cosmological constraints are also negligible.

\begin{figure}[!htbp]
    \includegraphics[width=3.4in]{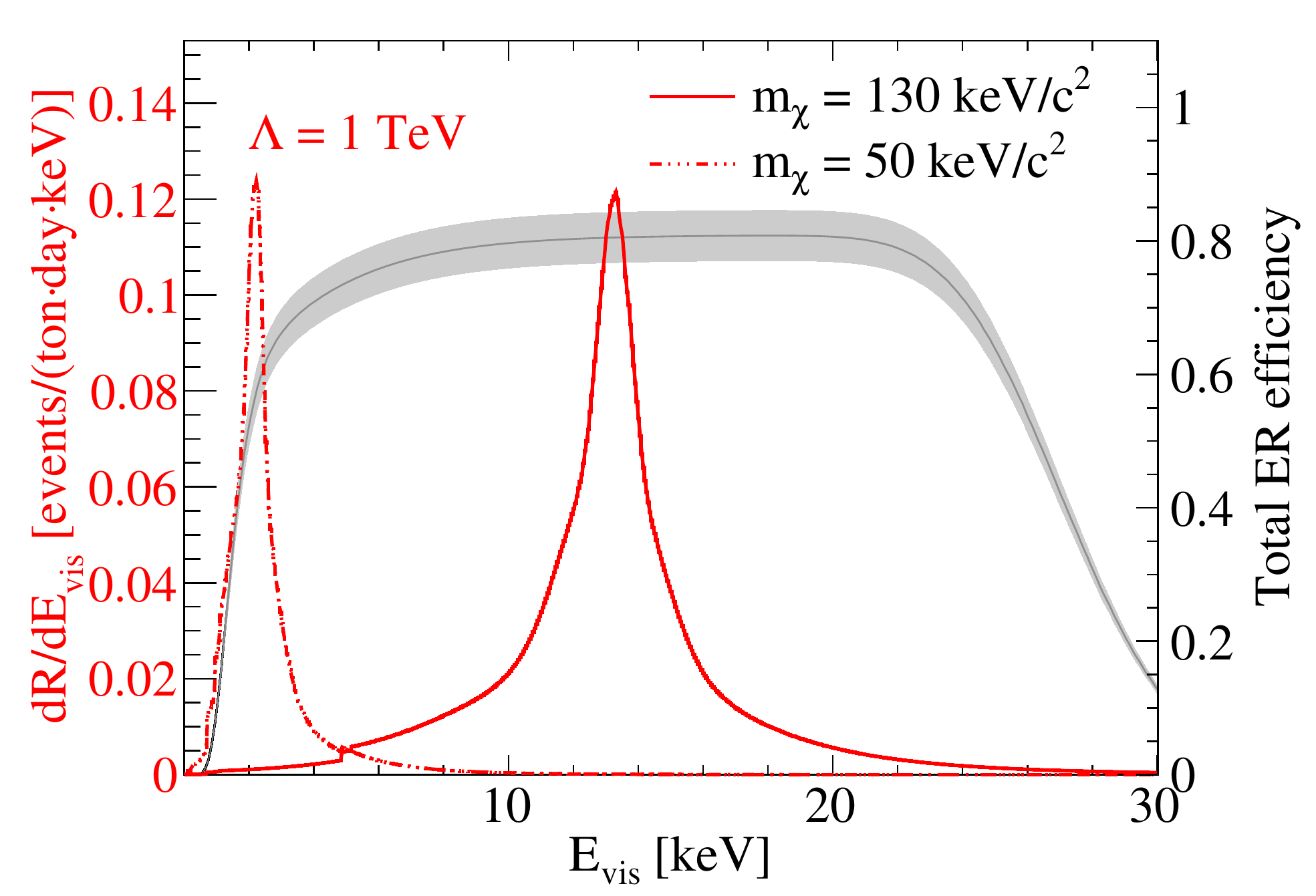}
    \caption{
    Visible energy spectra of the fermionic DM absorption on electron targets via a vector interaction for $m_{\chi} = 50$~keV/c$^2$ (dashed red line) and $m_{\chi} = 130$~keV/c$^2$ (solid red line) with $\Lambda = 1$~TeV, overlaid with detection efficiency and its error band with scale indicated on the right axis.
    }
    \label{fig:effAndDrde}
\end{figure}

Given the tiny mass of the active neutrino, the absorption signal has distinguishable peak features in the electron energy spectrum. Neglecting the halo velocity of the DM, 
we have
\begin{equation}
    m_{\chi} - E_{nl} = |\textbf{q}| + E_{R},
\end{equation}
where $m_{\chi}$ is the static mass of $\chi$, $E_{nl}$ ($E_{nl}>0$) is the binding energy of the initial electron on the state $|nl\rangle$, $|\textbf{q}|$ is the outgoing neutrino energy and $E_{R}$ is the recoil energy of the electron. If the binding energy is omitted, $E_{R}={m_{\chi}^2}/{2(m_{\chi}+m_e)}$.
The expected event spectrum of the visible energy $E_{\rm vis}$ summing up electron recoiling and deexcitation photon energy is
\begin{widetext}
\begin{equation}\label{eq:drde}
 \frac{dR}{dE_{\rm vis}} = N_{T}\frac{\rho_{\chi}}{m_{\chi}} \sum_{nl}(4l+2)\frac{|\textbf{q}|}{16\pi m_{e}^2 m_{\chi}(E_{\rm vis}-E_{nl})}|M(|\textbf{q}|)|^2 K_{nl}( (E_{\rm vis}-E_{nl}),|\textbf{q}|),
\end{equation}
\end{widetext}
where $N_T$ is the number of electron targets per unit mass, $m_e$ is the electron mass, and $\rho_{\chi}\sim 0.3$~GeV/cm$^{3}$/c$^2$ is the local DM density \cite{localDensity,plrWP}. 
$K_{nl}$ is the atomic $K$ factor~\cite{kfactor} also known as the ionization form factor~\cite{ytt,fFactor}. $|M(\textbf{q})|^2$ is the scattering amplitude with the leading term
\begin{equation}
\label{eq:sigma_def}
|M^{(V,A)}(|\textbf{q}|)|^2 = { (1,3)\times\frac{16\pi m_{e}^2 |\textbf{q}|}{m_{\chi}} (\sigma_{e}v_{\chi})}
\end{equation}
for $\mathcal{O}^{V}_{e\nu\chi}$ and $\mathcal{O}^{A}_{e\nu\chi}$, respectively. We define $\sigma_{e}\equiv{m_{\chi}^2}/{4\pi\Lambda^4}$ is the 
free-electron-$\chi$ total cross section for the vector operator when $m_{\chi}\ll m_{e}$.
Two examples of the event spectra with $m_{\chi}=50, 130$~keV/c$^2$ are shown in Fig.~\ref{fig:effAndDrde}. The peak will be further smeared by the actual detector resolution.

This work utilizes the data from the commissioning run (Run 1) of PandaX-4T~\cite{p4t}, which is located in the B2 Hall of China Jinping Underground Laboratory~\cite{cjpl,cjpl2}. The dual-phase liquid xenon time projection chamber (TPC) contains 3.7 tonne xenon in the sensitive region, viewed by two arrays of three-inch photomultipler tubes~(PMTs) from the top and the bottom. A recoiling event generates the prompt scintillation photons ($S1$) and the delayed electroluminenscence photons ($S2$). We use the same dataset as in the search for spin-independent WIMP-nucleus scattering~\cite{p4t}, with a fiducial volume~(FV) $2.67\pm 0.05$~tonne xenon and a 0.63~tonne-year exposure. 
The Run 1 data are divided into 5 subsets with slightly different field configurations and background levels. 
 For the range of the electron energy considered in this Letter ($< 30$~keV), the relativistic correction to the ionization form factor is less than 10\%, posing a minor effect on our search results.

The electron-equivalent energy of each event is reconstructed as
\begin{equation}
\label{e_rec}
    E_{\rm rec} = { 0.0137~{{\rm keV}}\times({\frac{ S1}{\rm PDE} + \frac{ S2_{b}}{ {\rm EEE}\times {\rm SEG}_{b} }})},
\end{equation}
where 0.0137~keV is the average work function of liquid xenon~\cite{workFunction}, PDE is the photon detection efficiency, {\rm EEE} is the electron extraction efficiency, and ${\rm SEG}_{b}$ is
 the single electron gain, using $S2_{b}$ collected by the bottom PMT array. ${G2}_{b}$ is conventionally defined as $ {\rm EEE}\times {\rm SEG}_{b}$ for the correction in $S2_{b}$. These parameters are predetermined using monoenergetic electronic recoil~(ER) peaks at the energy range from 41.5 to 408 keV~\cite{p4t}.

\begin{figure}[!htbp]
    \includegraphics[width=3.4in]{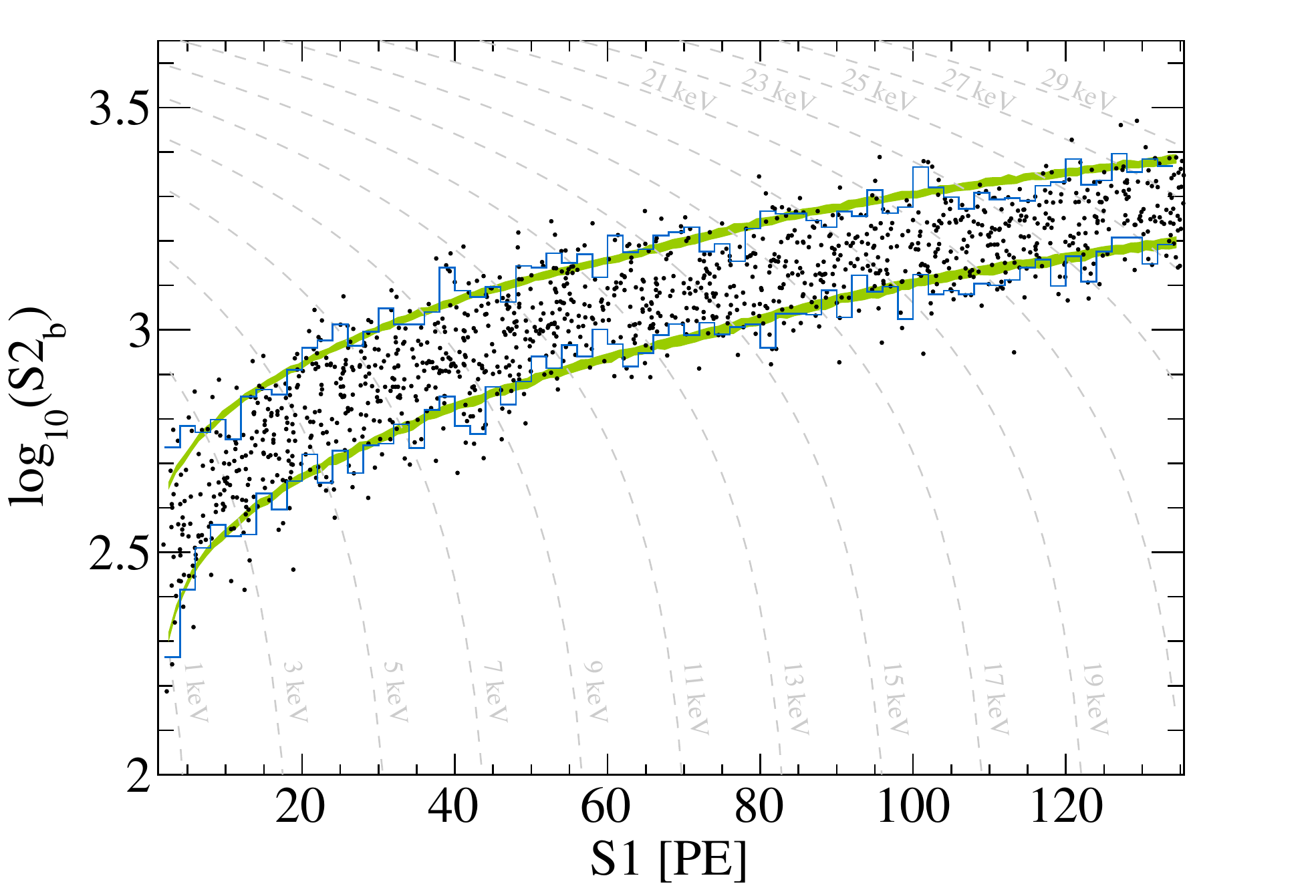}
    \caption{$^{220}$Rn calibration data in $\log_{10}(S2_b)$ vs. $S1$, taken under a drift field of 93~V/cm, and the comparison of 10\% and 90\% quantiles between the data (blue) and the best fit model (green). The average $1\sigma$ variation of $p_{*}$ in the quantiles is 9\%. The dashed gray lines are the equi-ER-energy lines. 
    }
        \label{fig:example} 
\end{figure}

Details of the ER signal response model will be discussed elsewhere~\cite{dzThesis}, and we will only outline steps relevant to this Letter. The detector ER response is calibrated by the 
$^{220}$Rn injection data (see Ref.~\cite{rn220} for the method). The model based on the noble element simulation techniques (NESTv2.2.1)~\cite{nest,nest2} is then used to fit the data below 30~keV, shown in Fig.~\ref{fig:example}. The so-called ionization recombination coefficient $r$, an intermediate random variable in the simulation, follows a Gaussian distribution with the median ($\mu_{\rm recomb}$) and the fluctuation ($\sigma_{\rm recomb}$). In the fit, $\mu_{\rm recomb}$ and $\sigma_{\rm recomb}$ are allowed to vary from nominal values from NEST by a quadratic function in $E_{\rm vis}$, and a scaling factor ($p_{f}$), respectively. Other detector parameters, PDE, ${G2}_{b}$, and SEG$_{ b}$, are treated as nuisance parameters.
After the fit, the quadratic parameters on $\mu_{\rm recomb}$
are linearly cast into three orthogonal variables via a principle component analysis, with the uncertainty dominated by the major parameter 
$p_0$ along the ``long axis" (30\% uncertainty). The fitted model agrees well with the distribution of calibration data.
The complete set of fitted detector parameters $p_{*}$ are summarized in Table~\ref{tab:modelPar}.

Because of the peak feature, understanding the detector resolution of ER events is critical for this search. 
Strong validation comes from the overall agreement between the data and model quantiles along each 
equienergy line in Fig.~\ref{fig:example}. Further validation comes from the measured 1$\sigma$ fractional energy resolution of $^{83m}$Kr data (41.5~keV and 6.8\%), which agrees with the prediction of our model (7.0\%)~\cite{absorptionNR}. 
Approximately, our model leads to a 1$\sigma$ resolution in $E_{\rm rec}$ as $0.073+0.173 E_{\rm vis}-6.5\times10^{-3}E_{\rm vis}^2+1.1\times10^{-4}E_{\rm vis}^{3}$ where $E_{\rm vis}$ is in unit of keV, and the smeared spectrum should be further corrected by the efficiency function in Fig.~\ref{fig:effAndDrde}. 
In our later analysis, however, the fit is performed in $S1$ and $S2_b$ using the full model, including uncertainties in $p_{*}$.

The main backgrounds in the light fermionic DM search follow the same compositions as in the WIMP search~\cite{p4t}, summarized in Table.~\ref{tab:modelPar}. The ER background components include tritium, flat ER (including radon, $^{85}$Kr, material background, and solar neutrino), $^{127}$Xe $L$-shell electron capture (5.2~keV), and $^{136}$Xe two-neutrino beta decay. The accidental, surface, neutron, and coherent solar neutrino-nucleus backgrounds are much less important in this ER analysis. 

\begin{table*}
    \caption{Summary of nominal values, uncertainties (fractional), and best fits for the detector response parameters $p_*$ (upper), and signal and individual background components (lower). Parameters $p_{*}$ include PDE, $G2_{b}$, SEG$_{b}$ and ionization recombination parameters $p_0$ and $p_{f}$ (see text for details). 
    Similar to Ref.~\cite{p4t}, the common DM signal and tritium background for each set are floating in the fit. The best fit values of the number of events have been corrected for their efficiencies. The change of the tritium rate between set 4 and 5 is attributed to the removal effects from the hot purifier~\cite{getter}. The variation in radon rate is primarily introduced from the krypton distillation tower.
    }
    \label{tab:modelPar}
\begin{ruledtabular}
    \centering
    \begin{tabular}{c c c c}
       Name  & Center of $p_{*}$ & $\sigma_{p{*}} $ & Best fitted $\delta_{p{*}}$ with DM data \\\hline
       
        PDE~[PE] &   0.0896 & 2\% & $(-0.4\pm1.7)\%$ \\
        $G2_{b}$ set 1, 2~[PE]& 3.5 &\multirow{2}{*}{6\% }  & \multirow{2}{*}{$(-0.8\pm1.2)\%$ } \\
        $G2_{b}$ set 3, 4, 5~[PE] & 4.3 & &  \\
        SEG$_{b}$ set 1, 2~[PE] & 3.8 & \multirow{2}{*}{2\% }  & \multirow{2}{*}{$(-1.1\pm 0.7)\%$ } \\
        SEG$_{b}$ set 3, 4, 5~[PE] & 4.6 & & \\
        $p_{0}$ & 0.124 & 30\% & $(9\pm29)\%$\\
        $p_{f}$ & 1.05 & 4\% & $(-4\pm 3)\%$ \\\hline
        Name  & $N_{b}\epsilon_{b}$  & $\sigma_{b}$ & Best fitted observed event number\\\hline
        Signal ($m_{\chi}=130$~keV/c$^2$) & float & N/A & {$47\pm23$}\\
        Tritium set 1 & float & N/A &{$16\pm 4$}\\
        Tritium set 2 & float & N/A & {$84\pm11$}\\
        Tritium set 3 & float & N/A & {$19\pm6$}\\
        Tritium set 4 & float & N/A & {$249\pm21$}\\
        Tritium set 5 & float & N/A & {$139\pm17$}\\
        Flat ER set 1, 2, 3, 5 & {\color{black}251.6} & 9\% & $242\pm16$\\
        Flat ER set 4 & {\color{black}240.5} & 9\% & $219\pm 15$ \\
        $^{136}$Xe & {\color{black}31.1} & 16\% & $32\pm5$\\
        $^{127}$Xe (L-shell electron capture) & {\color{black}8.13}  & 25\% & $8.5\pm2.0$ \\
        Accidental & 2.43 & 20\% & $2.4\pm0.5$\\
        Surface & 0.47 & 25\% & $0.44\pm0.11$\\
        Neutron & 1.15 & 50\% & $1.4\pm 0.6$\\
        $^{8}$B & 0.64 & 28\% & $0.60\pm0.17$\\\hline
        Total & & & $1060\pm46$\\
        Data & & & 1058\\
    \end{tabular}
\end{ruledtabular}

\end{table*}

The candidate signals are the same 1058 events as in Ref.~\cite{p4t}. 
To carry out the search, the data are fitted with a profile likelihood ratio analysis~\cite{Cowan:2010js}. 
The likelihood function for a given signal value $\mu$, evaluated in $S1$ and $S2_b$, is constructed as
\begin{widetext}
\begin{eqnarray}
\label{eq:likelihood}
  \mathcal{L}_{\rm tot} (\mu)=&&\big[\prod_{n=1}^{\textrm{nset}}\mathcal{L}_{n}\big] \times \big[\prod_{b} {G}(\delta_b, \sigma_b)\big]  \times \big[\prod_{p_{*}} {G}(\delta_{p_{*}}, \sigma_{p_{*}})\big]\,,\\
  \mathcal{L}_{n} = &&{\rm Poiss}(N_{\rm meas}^n|N_{\rm fit}^n)\times 
  \Bigg[\displaystyle{\prod_{i=1}^{N_{\rm meas}^n}}\left(\frac{N_{\mu}^n\epsilon_{\mu}^{n} P_{\mu}^n(S1^i,S2_b^i|\{p_{*}\})}{N_{\rm fit}^n}\right. 
+\left.\displaystyle\sum_{b} \frac{N_{b}^n\epsilon_{b}^{n}(1+\delta_{b})P_{b}^n(S1^i,S2_b^i|\{p_{*}\})}{N_{\rm fit}^n}\right)\Bigg].
\,,
\nonumber\\
{N_{\rm fit}^n}=&&N_{\mu}^n\epsilon_{\mu}^{n} + \sum_{b}{N_{b}^{n}\epsilon_{b}^{n}(1+\delta_{b})},\nonumber
\end{eqnarray}
\end{widetext}
where $\mathcal{L}_{n}$ is the likelihood function for dataset $n$, and the common Gaussian penalty terms $G$s are related to 
 uncertainties of the background and response parameters, with corresponding nuisance parameters (1$\sigma$
uncertainties) $\delta_b$ and $\delta_{p_{*}}$ ($\sigma_b$ and $\sigma_{p_{*}}$), respectively (Table.~\ref{tab:modelPar}). 
The measured number of events for each set $N_{\rm meas}^n$ is compared to the expected event number ${N_{\rm fit}^n}$ with a Poisson distribution. In each dataset, ${N_{\rm fit}^n}$ is a sum of fitted signal ($N_{ \mu}^n\epsilon_{\mu}^{n}$) and background event numbers ($N_{b}\epsilon_{\mu}^{n}$), with corresponding probability density distribution function (PDF) $P_{b}^{n}$ and $P_{\mu}^n$ in $S1$ and $S2_b$. 
The test static, $q_{\mu}$, is computed as the difference of $-2\log\mathcal{L}_{\rm tot}$ between the test signal value $\mu$ and the best fit~\cite{Cowan:2010js}. A cross-check of the fit performed in one-dimensional reconstructed energy is made, and the results are consistent.

In the fit, signal and background PDFs need to be updated for each set of response parameters $p_{*}$, which is computational expensive. 
The overall efficiencies $\epsilon_{b,\mu}^{n}$ are also functions of $p_*$ for different event components.
To include the uncertainties in $p_{*}$, we utilize a Monte Carlo reweighting technique~\cite{reweightMC,qcdReweight,lhcReweight,lhcReweight2} that a pool with millions of simulation events generated and saved ahead of time together with intermediate random variables drawn from $p_{*}$. When $p_{*}$ change to a different set of values, a new weight for the saved event $k$ is computed as the 
product of probability density ratios of all related intermediate variables. For example, if parameters $(p_{0},p_{f})$ are updated to $(p_{0}',p_{f}')$, the weight factor for the ionization recombination coefficient $r_k$ for the event $k$ is computed as
$\frac{G\textbf{(}r_{k}-\mu_{\rm recomb}(p_{0}'),\sigma_{\rm recomb}(p_{f}')\textbf{)}}{G\textbf{(}r_{k}-\mu_{\rm recomb}(p_{0}), \sigma_{\rm recomb}(p_{f})\textbf{)}}$, 
where $G$ again represents the Gaussian function. In the likelihood maximization, the common pool events reduce the statistical fluctuations, and with the graphics processing units acceleration, the variations in the five detector parameters are handled successfully.


\begin{figure}[!htbp]
    \includegraphics[width=3.4in]{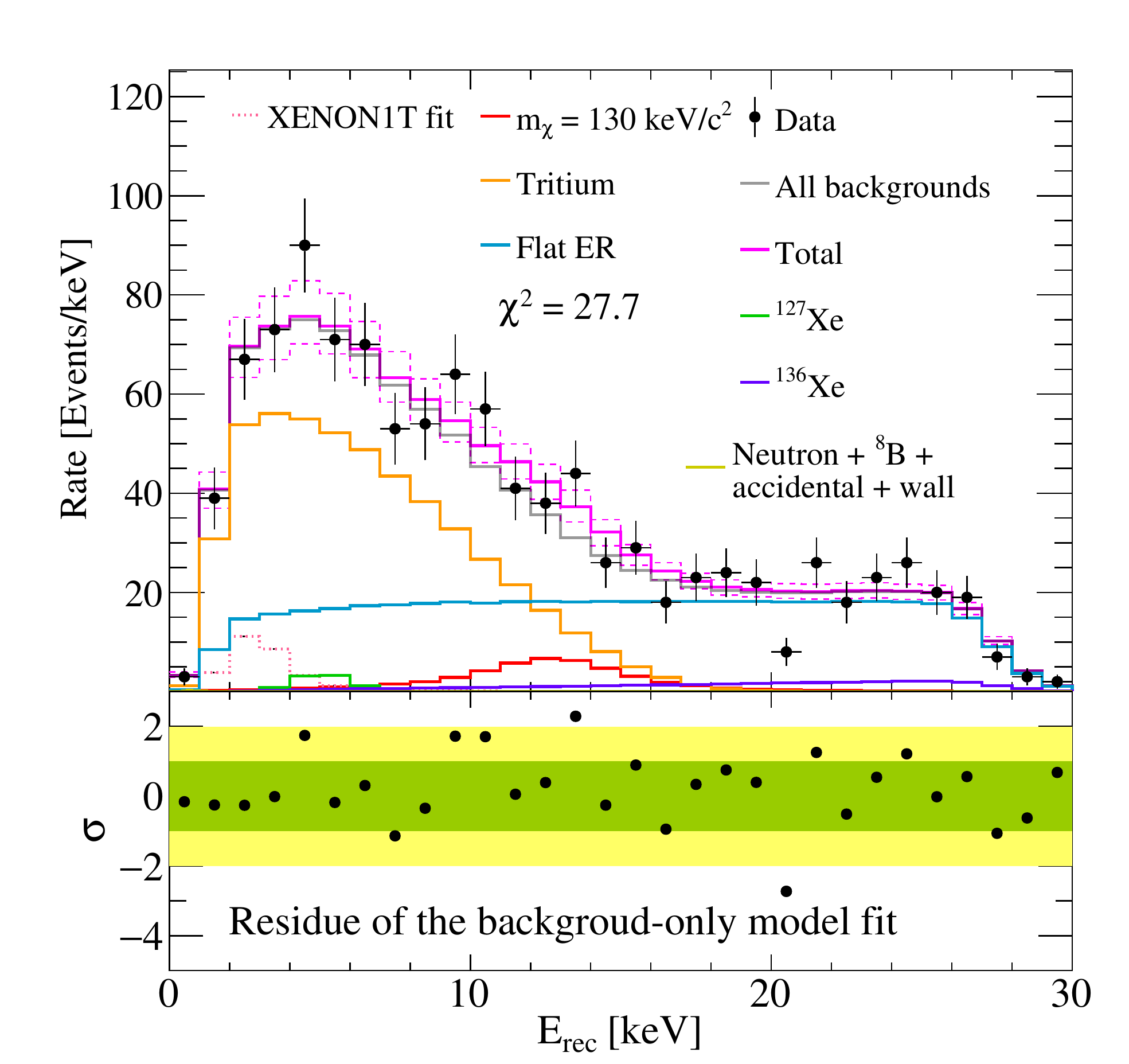}
    \caption{Top: The recoil energy spectrum $E_{\rm rec}$ data (black) overlaid with the best fit (solid magenta) with  
    $m_{\chi} = 130$~keV/c$^2$ (red histogram) and all background components as listed in Table.~\ref{tab:modelPar} (see legend). The gray histogram is the sum of the best fit background.  
    The dotted pink histogram is the best fit of XENON1T's low energy ER excess ($m_{\chi}=59$~keV/c$^{2}$ and $\Lambda=0.98$~TeV) from Ref.~\cite{gsf}. Bottom: the relative deviation of the data in the background-only model fit compared to the statistical uncertainty (green: $\pm1\sigma$, yellow: $\pm2\sigma$) in each 1~keV energy bin.
    }
    \label{fig:bestfit}
\end{figure}

The search of DM was carried out for $m_{\chi}$ larger than 10~keV/c$^{2}$ and smaller than {\color{black}180~keV/c$^{2}$}, well covered by the Run 1 data release~\cite{p4t}.
{\color{black} The global best fit is found at $m_{\chi} = 130$~keV/c$^2$, with all key components projected to $E_{\rm rec}$ (Eq.~\ref{e_rec}) in Fig.~\ref{fig:bestfit}, together with the best fit signal to XENON1T’s ER excess at $m_{\chi} = 60$~keV/c$^2$ (Ref.~\cite{gsf}) for comparison.
The computed $\chi^2$ using statistical uncertainties is 27.7 for 30~bins, indicating excellent agreement between the data and fit. 
The best fit detector and background nuisance parameters are all consistent with 0 within 2$\sigma$ (Table~\ref{tab:modelPar}). 
For the best fit signal strength, the local significance at $m_{\chi} = 130$~keV/c$^2$ is $1.7\sigma$ relative to zero, based on the $p$ value 
of $q_0$ (data tested with background-only hypothesis) evaluated using pesudodata generated with background-only simulations.
The significance reduces to $0.6\sigma$ with the look-elsewhere effect considered~\cite{plrWP,lee}. 
Note that the signal peak shape is rather smeared due to atomic effects and detector resolution. For example, at $m_{\chi}=130$~keV/c$^2$, the 
full width at half maximum is 5.2~keV. 
The slight and gentle excess between 9 to 14 keV is picked up by the fit, but much less so for narrower features (or fluctuations) in the spectrum, for example, at 4~keV. The $\sim$3$\sigma$ data deficit between 20 and 21!keV was investigated and concluded to be most likely a background downward fluctuation. 
}

\begin{figure}[!htbp]
     \includegraphics[width=3.4in]{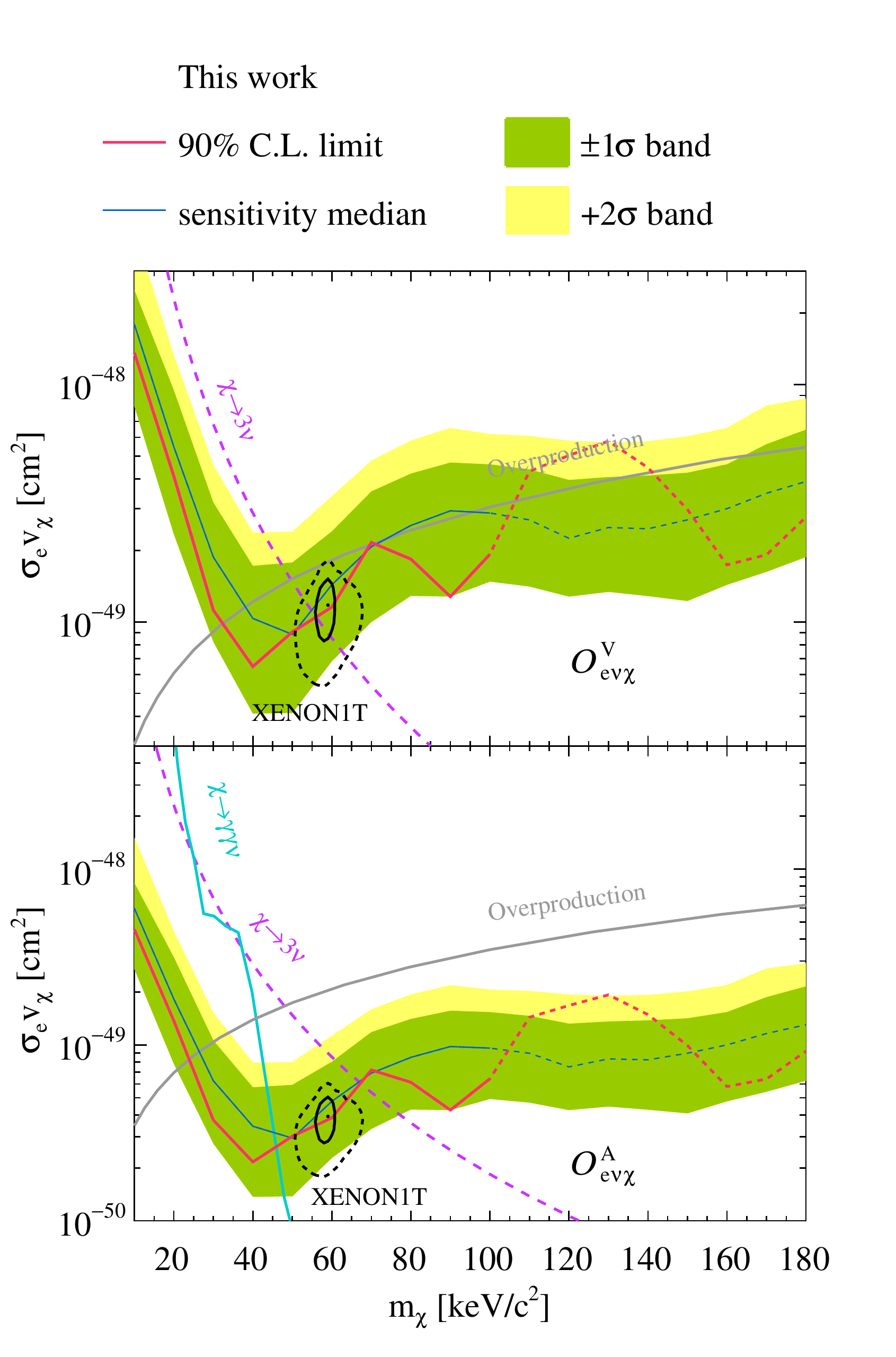}
     \caption{The 90\% CL exclusion limits (red lines) and $\pm$1 and 2$\sigma$ sensitivity (green and yellow band) on $\sigma_e v_{\chi}$ of fermionic DM absorption on electrons with the PandaX-4T commissioning data for the vector (upper) and axial-vector (lower) operators. Upper limits from the leading visible decays from x-ray satellites (cyan) whereas the leading process $\chi\to\gamma\gamma\gamma\nu$ for the vector operator is not shown which is above the parameter space selected, and cosmological constraints from leading invisible decays (dashed magenta) and DM overproduction (gray), and the best fit together with $1\sigma$ and $2\sigma$ contours associated with with 0.65 tonne$\cdot$year XENON1T data (black)~{\cite{gsf}} are overlaid.}
     \label{fig:limitSen}
 \end{figure}

With no significant excess identified, the 90\% confidence level (CL) upper limits on the $\chi e\to e\nu$ for the axial-vector and vector operators are set with the standard profile likelihood ratio analysis procedure, with background downward fluctuation power-constrained to $-1\sigma$~\cite{plrWP,powerC}.
The limit on axial-vector interaction is three-fold stronger than the vector interaction (Eq.~\ref{eq:sigma_def}).
The behavior of the sensitivity as a function of $m_{\chi}$ (blue line in Fig.~\ref{fig:limitSen}) is driven by several factors. 
According to Eqs.~\ref{eq:drde} and \ref{eq:sigma_def}, for a constant $\sigma_{e}v_{\chi}$, the scattering rate between DM and free electrons scales with $1/m_{\chi}^3$. 
In addition, both the detector threshold and background level contribute. For $m_{\chi}$ less than 20~keV/c$^{2}$, the sensitivity is worsened due to the threshold, whereas in the region where $m_{\chi}>90$~keV/c$^{2}$, the sensitivity flattens due to the reduction of background rate for $E_{\rm rec} \gtrsim 6$~keV (Fig.~\ref{fig:bestfit}). 
Note that for $m_{\chi}>100$~keV/c$^2$, the ionization form factor of the nonrelativistic atomic response contains non-negligible uncertainty, so we plot the limits in dashed lines~\cite{relativisticFactor}. The local upward fluctuation at $m_{\chi}=130$~keV/c$^2$, and the downward structures at $m_{\chi}=$ 90 and {\color{black}160~keV/c$^2$}, correspond to local data fluctuations around {\color{black} 13, 7, and 20~keV} in Fig.~\ref{fig:bestfit}, respectively.
For comparison, the astrophysical constraint from the leading visible decay channel ($\chi\to\gamma\gamma\nu$) for $\mathcal{O}^{A}_{e\nu\chi}$ 
from Ref.~\cite{gsf} is overlaid on Fig.~\ref{fig:limitSen} (blue line) combining results from Insight-HXMT~\cite{insight}, NuSTAR/M31~\cite{nustarM31} and INTEGRAL/08~\cite{integral08}, together with the cosmological constraint (dashed magenta) and the freeze-in overproduction constraint (orange){\color{black}~\cite{gsf}}. 
Our dark matter direct detection data have produced leading limits in the mass region 35 to 55~(25 to 45)~keV/c$^2$ for the vector~(axial-vector) operator. 
The $1\sigma$ amd $2\sigma$ contours obtained by fitting XENON1T data from Ref. \cite{gsf} are also overlaid.  Our limit is touching the center region of XENON1T’s best fit.

To summarize, we present the first sensitive experiment search on the absorption signals of fermionic dark matter on electron targets with an outgoing active neutrino with PandaX-4T. No significant dark matter signals are identified from the data.
Our data present the strongest constraints on the fermionic DM in the mass range 35 to 55~(25 to 45)~keV/c$^2$ for the vector operator (axial-vector operator), in comparison to constraints from x-ray satellites and large scale observations, illustrating the strong physics potential of PandaX-4T. The best fit to the XENON1T excess with the same model lies marginally within our constraint, which calls for further investigation with more data~\cite{lzprojected,xenon_nt,darkside20k,p4tProjected}.

We thank Tien-Tien Yu for the discussion on the ionization form factor, and Kenny C. Y. Ng and Brandon M. Roach for the discussion on the keV sterile neutrino. This project is supported in part by a grant from the Ministry of Science and Technology of China (No. 2016YFA0400301), grants from National Science Foundation of China (Nos. 12090061, 12090064, 12005131, 11905128, 11925502, 11775141), and by Office of Science and Technology, Shanghai Municipal Government (grant No. 18JC1410200). This project is also funded by China Postdoctoral Science Foundation (No. 2021M702148). We thank supports from Double First Class Plan of the Shanghai Jiao Tong University. We also thank the sponsorship from the Chinese Academy of Sciences Center for Excellence in Particle Physics (CCEPP), Hongwen Foundation in Hong Kong, and Tencent Foundation in China.  Finally, we thank the CJPL administration and the Yalong River Hydropower Development Company Ltd. for indispensable logistical support and other help.

 $Note$ $added$.----Recently, the XENONnT Collaboration released a new dataset~\cite{xenonnTnew}, excluding XENON1T’s excess. The new data are also consistent with our constraints.








\bibliographystyle{apsrev4-1}
\bibliography{absorption.bib}

\end{document}